# Where Density Functional Theory Goes Wrong and How to Fix it: Spin Balanced Unrestricted Kohn-Sham Formalism

Artëm Masunov

*Theoretical Division,T-12, Los Alamos National Laboratory, Mail Stop B268, Los Alamos, NM 87545.*

**Submitted October 22, 2003**; amasunov@lanl.gov

**ABSTRACT**: Kohn-Sham (KS) formalism of Density Functional Theory is modified to include the systems with strong non-dynamic electron correlation. Unlike in extended KS and broken symmetry unrestricted KS formalisms, cases of both singlet-triplet and *aufbau* instabilities are covered, while preserving a pure spin-state. The straightforward implementation is suggested, which consists of placing spin constraints on complex unrestricted Hartree-Fock wave function. Alternative approximate approach consists of using the perfect pairing implementation with the natural orbitals of unrestricted KS method and square roots of their occupation numbers as configuration weights without optimization, followed by *a posteriori* exchange-correlation correction. The numerical results of this approximation for the barrier to the internal rotation in ethylene are reported to be in close agreement with experimental data.

Density Functional Theory (DFT) is becoming a widely used tool in theoretical chemistry. Restricted Kohn-Sham (RKS) formalism of DFT,[1] based on single Slater determinant description for noninteracting system, is implemented nowadays in most standard quantum chemistry packages. For many molecular systems its results are close in accuracy to those of coupled cluster method at computational cost nearly equal to the one of Hartree-Fock (HF) method. This accuracy, however, is less consistent for "difficult" molecular systems which require several determinants for their description in molecular orbital (MO) theory.[2] The present Communication is aimed to investigate this inconsistency.

The method presently known as RKS was first introduced by Slater in 1951 as a simplification to the HF method.[3] Two-electron Fock matrix in HF formalism contains two distinct terms: Coulomb $<ij|1/r|ij>$, originating form electron-electron repulsion, and exchange $<ij|1/r|ji>$, which arise from antisymmetric form of HF wavefunction (Slater determinant). While Coulomb term has classical interpretation of electrostatic interaction between electron density distributions on MOs $>i|i<$, exchange term has no classic equivalent and may be interpreted as self-interaction of transition densities $>i|j<$ (differential overlap between two MOs). Slater suggested replacing the exchange term in HF equations (the most computationally expensive part) with the approximate expression for uniform electron gas, which nonlinearly depends on electron density. This approach was formalized in 1965 by Kohn and Sham,[1] who showed that it is in principle exact, if one uses exact exchange-correlation (XC) potential instead of Slater exchange. This XC potential is defined as an external potential necessary to keep the total electron density of a hypothetical system consisting of non-interacting electrons equal to the exact electron density of the real physical system.

The functional form of this XC potential remains unknown to the present day. However, numerous approximations had been suggested,[4] the remarkable accuracy of KS calculations results from this long quest for better XC functional. One may classify these approximations into local (depending on electron density, or spin density), semilocal (including gradient corrections), and nonlocal (orbital dependent functionals). In this classification HF method is just one of non-local XC functionals, treating exchange exactly and completely neglecting electron correlation.

First and the most obvious reason for RKS performance inconsistency for the "difficult" systems, mentioned above, is imperfection of the approximate XC functionals. Second, and less known reason is the fact that KS approach is no longer valid if electron density is not *v*-representable,[5] i.e., there is no ground state single Slater determinant for a given XC potential. Examples of such densities (obtained from high level MO calculations and thus essentially exact) were found by Baerends in molecules $CH_2$ and $C_2$.[6] To include these important cases, Ullrich and Kohn developed extended KS (EKS) formalism, which describes non-interacting system by an ensemble (linear combination) of several determinants, which differ form one another with just one orbital. Equivalent (and more convenient for practical implementations[7]) is the description which retains single determinant, but allows for partial occupation numbers of several degenerate orbitals.

Preliminary EKS calculations indicate that the result deviates from RKS only in the close vicinity of degeneracy,[8] and an approximate XC functional makes this range even smaller.[9] Comparison of the potential curves for intramolecular twist in ethylene[8] showed that EKS curve is closer to (incorrect) RKS one, but unrestricted KS (UKS) curve is closer to almost exact one, obtained with CAS. The reason for this EKS failure is that it addresses only one kind of instabilities (namely *aufbau* instability) observed in calculations,[10] and studied theoretically.[11] The *aufbau* instability is observed when the energy of the lowest unoccupied MO (LUMO) is less then that of highest occupied MO (HOMO). The attempt to replace HOMO with LUMO in single determinant description raises the energy of the former and lowers the energy of the latter, thus retaining *aufbau* instability.

There is also another kind of instability of RKS determinant, called singlet-triplet instability. It can be avoided if different spatial orbitals are used for α and β spin subsystems (spin polarized or broken spin symmetry solution, BS). BS UKS description has lower total energy, but predicts unphysical spin polarization and nonzero spin density ($\rho_\alpha$-$\rho_\beta$) for closed shell systems. Advantages and disadvantages of UKS vs. RKS approaches from the practical standpoint were recently reviewed by Cremer.[12] From the formal theory standpoint BS UKS was addressed by Perdew, Savin, and Burke[13] in a rather paradoxal manner. They postulated that UKS gives the correct total density, while the spin density is incorrect, and suggested considering on-top pair electron density instead.

For a long time BS UHF wavefunction was known not to be an eigenfunction of the spin operator $S^2$. Its value (taken as a measure of spin contamination) is typically intermediate between correct singlet and triplet values (or, in extreme cases, higher multiplets). Geometry optimization with BS UHF usually results in a structure, intermediate between singlet and triplet. The reason for this and other disadvantages of BS description is easy to see in the case of two-electron open shell system, where the electrons occupy orthogonal spatial orbitals $i$ and $a$ (HOMO and LUMO). Correct singlet wavefunction of this system requires two determinants: $(|i\alpha a\beta> - |i\beta a\alpha>)/\sqrt{2}$, while triplet can be described by tree wavefunctions (degenerate in the absence of external field): $(|i\alpha a\beta> + |i\beta a\alpha>)/\sqrt{2}$, $|i\alpha a\alpha>$ and $|i\beta a\beta>$. Thus, the single determinant of UHF $|i\alpha a\beta>$ is an average between the singlet and the first one of triplet wavefunctions (50/50 spin contamination).

In the present contribution I propose to avoid spin contamination in UKS by adding a second determinant, where spatial parts of α and β sets are interchanged. Such two-determinant UKS shall be called spin-balanced UKS (SB UKS). It is easy to see that SB UKS, similar to EKS description, is equivalent to RKS wherever RKS solution is stable. This is not the case for multiconfigurational approaches, derived from KS method (like CAS-DFT[14] or CI-DF[15]). Unlike EKS, SB UKS approach covers both *aufbau* and singlet-triplet instabilities, since removing the latter eliminate the former as well. Thus, SB UKS is more general than EKS.

The practical implementation of SB UKS is straightforward within complex UHF formalism:[16] single determinant built on complex orbitals has real and complex parts (and thus, is equivalent to the two-determinant wave function). Complex KS orbitals must be optimized self-consistently with one additional constraint: imaginary part of each spatial orbital in alpha set is equal to the real part of the spatial orbital from the beta set with the same quantum number. The total electron density of such spin-coupled two-determinant wave function is equivalent to the one of a single determinant, but spin density and total energy differ. Thus, Perdew, Savin, and Burke's postulate[13] becomes a theorem and spin symmetry dilemma is resolved in a more elegant way. One may speculate that optimized orbitals in SB wavefunction will be close to MOs from BS solution when it exists and to MOs of restricted method when spin-polarization is negligible. But unlike BS, spin polarization of the SB orbitals will occur when two-determinant description is needed, not only when there is a low in energy triplet state.

Similar to other multiconfigurational approaches in DFT,[14,15] double counting of electronic correlation needs to be addressed. In order to do that, it is useful to separate the electronic correlation into two parts: dynamic (short range) and non-dynamic (long range, arising from interactions of nearly-degenerate states). Dynamic correlation in molecules is similar to the one in the free electron gas, and thus is well described by the correlation part of the XC functionals. Nondynamic correlation is simulated by the exchange part, when the nonlocal exact HF exchange is replaced with an approximate local functional.[12] Specifically, it is self interaction error (exchange interaction of the KS orbital with itself) that helps to describe the non-dynamic correlation and contributes to the success of the RKS method. SB UKS approach explicitly describes both exchange and dynamic correlation. The active space in SB UKS includes all the orbitals, which eliminates the need in the elaborate separation of the electron density into correlated and uncorrelated parts.[14] On the other hand, minimalist two-determinant description ensures that no dynamic correlation is included explicitly. For this reason only the correlation part of XC functional needs to be included in SB

UKS calculations. Similar reasoning was applied recently[17] to correct the energy of mutideterminant wavefunction for dynamic correlation.

The existing implementations of complex UHF are not well suited to impose SB constraint. For this reason in the present contribution the implementation of the perfect pairing (PP) scheme[18] (equivalent to complex UHF) was used. To demonstrate their equivalence, one may use natural orbitals (NOs) of UHF approach. NOs are eigenvectors of the total density matrix. They come in pairs φ, ψ with occupation numbers of $\lambda^2$ and $2-\lambda^2$ (λ=1 for open-shell orbitals, and fractional otherwise). In the following example of twisted ethylene only one pair of fractionally occupied NOs is found. The rest is (nearly) doubly occupied and form inactive core $\Phi_D$. Detailed analysis shows[19] that BS UHF orbitals $\chi^\alpha$, $\chi^\beta$ can be expressed using NOs: $\chi^\alpha=\mu\varphi+\lambda\psi$ and $\chi^\beta=\mu\varphi-\lambda\psi$, where $2\mu^2=1+S$ and $2\lambda^2=1-S$, and $S=<\chi^\alpha|\chi^\beta>$. Typically overlap $S$ is large, so that the spin-polarization parameter λ is small and μ is close to unity. In BS UHF description the wavefunction is $\Psi=|\Phi_D\chi^\alpha\alpha\chi^\beta\beta>=\mu^2|\Phi_D\varphi\alpha\varphi\beta>-\lambda^2|\Phi_D\psi\alpha\psi\beta>+\sqrt{2}|\Phi_D\varphi\psi(\alpha\beta+\beta\alpha)>$, where the last term represents the triplet component (spin contamination). In SB description above, the wavefunction is $\Psi=(|\Phi_D\chi^\alpha\alpha\chi^\beta\beta>-|\Phi_D\chi^\alpha\beta\chi^\beta\alpha>)/\sqrt{2}=\mu^2|\Phi_D\varphi\alpha\varphi\beta>-\lambda^2|\Phi_D\psi\alpha\psi\beta>$, thus spin contamination is cancelled, but NOs and occupation numbers $\mu^2$ and $\lambda^2$ remain the same. This description is readily generalized for arbitrary number of fractionally occupied NOs. Wavefunction of this exact form is used in PP method, except that CI coefficients ($\mu/\sqrt{2}$ and $\lambda/\sqrt{2}$) and NOs (φ and ψ), are optimized.

Here I report the results on internal rotation barrier in ethylene. In all the calculations Gaussian 98 suite of programs[20] was used. The BS unrestricted solution was obtained, NOs and occupation numbers were used in PP calculation unchanged (denoted PPNO in the following).[21] For the sake of comparison with Ref.8, the same molecular geometry, basis set (4-31G) and XC functional (SVWN5) were used.

The rotation barrier heights are compared with EKS[8] and experimental[22] results in **Table 1**. The restricted approach (RHF and RKS) does not work well. EKS approach leads to only marginal improvement. Unrestricted approach (UHF and UKS) works better at the expense of spin contamination. PP approach using UHF results (PPNO-UHF) describes pure singlet, and gives the barrier value close to CAS. PP approach using BS UKS (HF+VWN5) orbitals underestimates the barrier. After the correlation correction (defined as the difference between UKS and UHF energies, evaluated with UKS orbitals) is added, the result gives the best agreement with the experiment.

**Table 1.** Total energy $E_{tot}$ (hartree) and relative barrier $E_b$ (kcal/mol) for internal rotation in ethylene.

| Method | $E_{tot}$ | $E_b$ | error |
|---|---|---|---|
| Experiment (ref.21) |  | 65.0 | 0.0 |
| UCCSD | -78.1303 | 72.1 | 7.1 |
| CASMP2(2,2) | -78.0957 | 72.8 | 7.8 |
| CAS(2,2) | -77.9515 | 74.5 | 9.5 |
| RHF | -77.9204 | 112.6 | 47.6 |
| UHF | -77.9206 | 47.1 | -17.9 |
| RKS | -77.7214 | 97.3 | 32.3 |
| EKS (ref.8) |  | 95.9 | 30.9 |
| UKS | -77.7214 | 77.9 | 12.9 |
| PPNO-UHF | -77.9461 | 71.3 | 6.3 |
| PPNO-UKS | -77.9196 | 55.3 | -9.7 |
| PPNO-UKS+C | -78.9068 | 63.3 | -1.7 |

When combined with the exact dynamic correlation operator, which was developed recently from the first principles,[23] SB UKS approach holds a promise to be the final word in half a century long quest for the exact DFT.

In conclusion, this work introduces the new modification of DFT method, SB UKS. It addresses both singlet-triplet and *aufbau* instabilities of restricted single determinant description. The new method can be implemented using constrained complex UKS orbitals. Alternatively, a close approximation can be used where broken symmetry UKS solution exists. It employs existing perfect pairing scheme, but NOs and their occupations are taken from UKS results instead of variational procedure. The results of this approximation are reported for the barrier to internal rotation in ethylene. They are found to be in a close agreement with experiment.

**Acknowledgments.** Encouraging and educational discussions with Dr. Richard L. Martin and Prof. Vitaly A. Rassolov are gratefully acknowledged. This work was supported in part by LDRD program at LANL, LA-UR-03-7824.